\documentclass{fic-l}
\usepackage{amssymb}
\renewcommand{\theequation}{\thesection.\arabic{equation}}
 \newcommand{\begeq}{\begin{equation}}
\newcommand{\bea}{\begin{eqnarray}}
\newcommand{\eea}{\end{eqnarray}}

\newcommand{\ca}{$C^*$-algebra} 
 \newcommand{\rep}{representation}

\newcommand{\Hs}{Hilbert space}

   \newcommand{\vna}{von
Neumann algebra}
\newcommand{\op}{^{\mbox{\tiny op}}}
\newcommand{\id}{\mbox{\rm id}}

 \newcommand{\ovl}{\overline}
 
\newcommand{\raw}{\rightarrow} 
\newcommand{\law}{\leftarrow} \newcommand{\Raw}{\Rightarrow}
 \newcommand{\Law}{\Leftarrow}

\newcommand{\LRaw}{\Leftrightarrow}
\newcommand{\rlh}{\rightleftharpoons} \newcommand{\n}{\|}
\newcommand{\ot}{\otimes} 
\newcommand{\la}{\langle} \newcommand{\ra}{\rangle}

\newcommand{\x}{\times} 
 
\newcommand{\cin}{C^{\infty}}

\newcommand{\inv}{^{-1}}

\newcommand{\bb}{\rangle_{\mathfrak{B}}}

 \newcommand{\Gm}{\Gamma}

\newcommand{\io}{\iota} 
 
\newcommand{\rh}{\rho} \newcommand{\sg}{\sigma}
\newcommand{\Sg}{\Sigma} \newcommand{\ta}{\tau} 
\newcommand{\Ph}{\Phi} \newcommand{\phv}{\varphi}
  \newcommand{\Ps}{\Psi}
 
\newcommand{\A}{\mathfrak{A}} \newcommand{\B}{\mathfrak{B}}
\newcommand{\GC}{\mathfrak{C}}

\newcommand{\GM}{\mathfrak{M}} \newcommand{\GN}{\mathfrak{N}}

\newcommand{\GP}{\mathfrak{P}} 

 \newcommand{\CE}{{\mathcal E}}
 \renewcommand{\H}{{\mathcal H}}
 
\newcommand{\CK}{{\mathcal K}}   \newcommand{\CL}{{\mathcal L}}   
 \newcommand{\CM}{{\mathcal M}}

\renewcommand{\L}{{\mathcal L}}
\newcommand{\C}{{\mathbb C}}

  \makeatletter
\newskip\tempskip \def\endproof{{\parfillskip24\p@ plus\@ne
fil\@@par}\tempskip\prevdepth
\ifdim\lastskip=\z@\tempskip\z@\else\vskip-\lastskip
\ifdim\tempskip>4\p@ \tempskip.5\tempskip \else \tempskip\z@\fi\fi
\nobreak\vskip-\baselineskip\vskip-\tempskip\noindent\hbox
to\hsize{\hfill
$\blacksquare$}\par\vskip\tempskip\vskip\abovedisplayskip\@doendpe}
\makeatother \makeatletter
\newskip\tempskip \def\endiproof{{\parfillskip24\p@ plus\@ne
fil\@@par}\tempskip\prevdepth
\ifdim\lastskip=\z@\tempskip\z@\else\vskip-\lastskip
\ifdim\tempskip>4\p@ \tempskip.5\tempskip \else \tempskip\z@\fi\fi
\nobreak\vskip-\baselineskip\vskip-\tempskip\noindent\hbox
to\hsize{\hfill
$\Box$}\par\vskip\tempskip\vskip\abovedisplayskip\@doendpe}
\makeatother 

\newtheorem{theorem}{Theorem}[section]

\newtheorem{definition}[theorem]{Definition}

\newtheorem{proposition}[theorem]{Proposition}
\newtheorem{corollary}[theorem]{Corollary}

\theoremstyle{remark}
\newtheorem{remark}[theorem]{Remark}
\newtheorem{example}[theorem]{Example}

\numberwithin{equation}{section}

\newcommand{\otc}{\circledcirc}

\newcommand{\Me}{Morita equivalent}

\newcommand{\Rep}{\mbox{\rm Rep}}
\newcommand{\End}{\mbox{\rm End}}
\newcommand{\Hom}{\mbox{\rm Hom}}
\newcommand{\Der}{\mbox{\rm Der}}

\newcommand{\Ri}{\mbox{\rm [{\sf Rings}]}}
\newcommand{\Ca}{\mbox{\rm [{\sf C}$\mbox{}^*$]}}
\newcommand{\Wa}{\mbox{\rm [{\sf W}$\mbox{}^*$]}}
\newcommand{\LG}{\mbox{\rm [{\sf LG}]}}
\newcommand{\SyG}{\mbox{\rm [{\sf SG}]}}
\newcommand{\Po}{\mbox{\rm [{\sf Poisson}]}}
\begin{document}

\title{Bicategories of operator algebras and Poisson manifolds}

\author{N.P. Landsman}
\address{Korteweg--de Vries Institute for Mathematics,
University of Amsterdam,\\
Plantage Muidergracht 24,
NL-1018 TV AMSTERDAM, THE NETHERLANDS}
\email{npl@science.uva.nl}
\thanks{Supported by a Fellowship from the Royal Netherlands Academy
of Arts and Sciences (KNAW)}
\thanks{
Dedicated to Sergio Doplicher and John Roberts, on their sixtieth birthday}
\date{\today}

\begin{abstract}
It is well known that rings are the objects of a bicategory, whose
 arrows are bimodules, composed through the 
bimodule tensor product. We give an analogous bicategorical
description of \ca s, von Neumann algebras, Lie groupoids, symplectic
groupoids, and Poisson manifolds.  The upshot is that known
definitions of  Morita equivalence for any of these cases
amount to isomorphism of objects in the pertinent bicategory.
\end{abstract}
\maketitle
\section{Introduction}
One of the achievements of Doplicher and Roberts has been to introduce
 monoidal categories into quantum field theory; indeed, their
work provides the most convincing example of categorical thinking in
physics by a long shot. Since a monoidal category is nothing
but a bicategory with one object, it seems an appropriate tribute to
describe a number of more general bicategories with proven relevance
to physics. 

In a bicategory \cite{Ben,MacLane} (alternatively called a weak 2-category,
which is a special case of a weak $n$-category \cite{BD}), 
each space of arrows between two fixed objects is
itself a category, so that there is a notion of isomorphisms of such arrows. 
To distinguish the arrows between objects in a bicategory from the arrows
in each category of arrows, the former arrows are called horizontal,
the latter vertical. Now, horizontal composition of arrows is
merely associative up to isomorphism, and the horizontal
 composition of 
an arrow with a local unit arrow is merely
isomorphic to the given arrow, rather than equal to it, as in a category.

Our examples of bicategories originated in the theory of constrained
quantization \cite{NPL3,NPLOBWF},
relating classical and quantum physics in
an intriguing way. 
In the order: Objects, horizontal arrows, composition operation, unit
arrows (as in: Rings, bimodules, algebraic bimodule tensor product,
canonical bimodule over a ring),
these examples are
\begin{enumerate}
\item
 \ca s, Hilbert ($C^*$) bimodules \cite{Rie}, Rieffel's tensor product
 \cite{Rie}, canonical Hilbert bimodule;
\item
 \vna s, correspondences \cite{Con}, Connes's tensor product
 \cite{Con,Sau}, standard form \cite{KR2};
\item
Lie groupoids \cite{Mac}, regular bibundles \cite{Hae},
Hilsum--Skandalis tensor product \cite{HS}, canonical bibundle;
\item
Symplectic groupoids \cite{CDW},  regular \cite{NPLOBWF} symplectic
bibundles \cite{Xu1},  Hilsum--Skandalis--Xu tensor product \cite{Xu1},  
canonical symplectic  bibundle;
\item
 Integrable Poisson manifolds \cite{CDW},
regular \cite{NPLOBWF} symplectic bimodules (dual pairs) \cite{W1,K1}, 
Xu's tensor product
\cite{Xu1,NPL3}, 
s-connected and s-simply connected symplectic groupoid.
\end{enumerate}

In the literature one finds notions of Morita equivalence for each
of these cases (see \cite{Rie,Rie,MRW,Xu1,Xu2}, respectively).
It is gratifying that all these definitions may be derived
from a single notion, namely isomorphism of objects in the pertinent
bicategory.

The plan is as follows: we first recall the notion of a bicategory,
then briefly discuss rings as a warmup, and
subsequently explain each of the above cases in some detail, including
the pertinent Morita theory.

\textbf{Acknowledgements} The author is indebted to 
M. M\"{u}ger and I. Moerdijk for drawing his attention to
bicategories.
\section{Bicategories}
Our notation for (bi)categories (including groupoids) will be that $C$
denotes a (bi)category as a whole, whose class of objects is $C_0$, and
whose class of arrows is $C_1$. For $a,b\in C_0$, the Hom-space
$(a,b)\subset C_1$ stands for the collection of arrows from $b$ to
$a$, so that composition of arrows is a map from $(a,b)\x (b,c)$ to
$(a,c)$. A functor $F:C\raw D$ decomposes as $F_0:C_0\raw
D_0$ and $F_1:C_1\raw D_1$, subject to the usual axioms.
The unit arrow associated to an object $a\in C_0$ is denoted $1_a\in (a,a)$.

Categories can sometimes be ``enriched'' so as to become 2-categories.
Although we will not encounter this precise structure, 2-categories are a 
special case of bicategories, which is easier to explain, and may serve
as a warmup for the latter. The following material comes straight from
\cite{MacLane}; also see \cite{BD} for an enthusiastic account, with
generalizations to $n$-categories.
\begin{definition}
A 2-category is a category $C$ for which each class of arrows
$(a,b)$, $a,b\in C_0$, is itself a category, such that
\begin{enumerate}
\item
The map $(a,b) \x (b,c)\raw (a,c)$ given by composition of arrows in
$C_1$ is the 0-component of a functor (where the left-hand side is the
Cartesian product of two categories; Mac Lane \cite{MacLane} then
speaks of a bifunctor);
\item
For each $a\in C_0$, the map $U_a:1\mapsto 1_a$ is the
0-component of a functor from the trivial (or terminal)
category $T$ (with one object
and one arrow) to $(a,a)$.
\end{enumerate}
\end{definition}

These axioms simply mean that the given maps can be extended to well-behaved
maps between ``arrows between arrows''. The first motivating example
of a 2-category is the category of all categories, with functors
as arrows, and natural transformations as arrows between arrows.
The second is the category of topological spaces, with continuous maps
as arrows, and homotopies as arrows between arrows.

In the context of this paper, the category of rings with homomorphisms
as arrows should be seen as a 2-category, with intertwiners as arrows
between arrows. \ca s and \vna s with (normal)
$\mbox{}^*$-homomorphisms, and Lie groupoids with smooth functors
provide further examples.

Bicategories generalize 2-categories, as follows.
\begin{definition}\label{defbi}
A bicategory (or weak 2-category)  $C$ 
consists of
\begin{itemize}
\item
A class of objects $C_0$;
\item
A category $(a,b)$ for each pair $(a,b)\in C_0\x C_0$;
\item
A (bi)functor $\Ph_{a,b,c}:(a,b) \x (b,c)\raw (a,c)$ for each triple $(a,b,c)$;
\item
A functor $U_a:T\raw (a,a)$, $U_a(1)=1_a$, for each $a\in C_0$, 
\end{itemize}
such that
\begin{enumerate}
\item The functors
$\Ph_{a,c,d}\Ph_{a,b,c}$ and $\Ph_{a,b,d}\Ph_{b,c,d}$ from $(a,b)\x
(b,c)\x (c,d)$ to $(a,d)$ are naturally isomorphic;
\item 
The  functors  
$f\mapsto f1_b$ (where $f\in (a,b)$) and $\id_{(a,b)}$
from $(a,b)$ to itself are naturally isomorphic;
\item The  functors  
 $1_af\mapsto f$ and $\id_{(a,b)}$
from $(a,b)$ to itself are naturally isomorphic,
\end{enumerate}
subject to coherence laws stated on p.\ 282 of \cite{MacLane}
(these laws lead to consistency of various orders of bracketing).
\end{definition}

A 2-category is a bicategory in which the natural isomorphisms in
the above definition are the identity.
Many examples of genuine bicategories will be given in what follows.
The bifunctors $\Ph$ are sometimes said to define the ``horizontal''
composition of arrows, in contradistinction to the ``vertical''
composition of arrows in each of the categories $(a,b)$.

Since the theory of Morita equivalence will involve isomorphism of
objects in a bicategory, we should point out that this notion is
weaker than in a category.
\begin{definition}\label{eqbi}
 Two objects $a,b$ in a bicategory are
isomorphic, written $a\cong b$, when there exist arrows $f\in (a,b)$
and $g\in (b,a)$ such that $fg\simeq 1_a$ (isomorphism in the usual
sense as objects in the category $(a,a)$) and $gf\simeq 1_b$ in
$(b,b)$.
\end{definition}
\section{Rings}
The following theorem is already mentioned (as an example) in
\cite{Ben,MacLane}. 
\begin{theorem}\label{biring}
For any two rings $R,S$, let $(R,S)$ be the collection of all bimodules
$R\raw M\law S$, seen as the class of objects of a category, whose
arrows are $R$-$S$ linear maps. 

The collection of all rings as
objects, bimodules as arrows,  (horizontal) composition $(R,S)\x
(S,T)\raw (R,T)$ given by $\ot_S$, and the unit arrow $1_S$ in $(S,S)$ given
by the canonical bimodule $S\raw S\law S$, is a bicategory \Ri.
\end{theorem}

One looks at bimodules as generalized homomorphisms. The precise connection
between bimodules and ordinary homomorphisms is as follows.
\begin{remark}\label{ringmor}
Given a (unital) homomorphism $\rh:R\raw S$, one constructs a
bimodule $R\raw S\law S$ by $r(s)=\rh(r)s$ and $(s)t=st$.
We write this as $R\stackrel{\rh}{\raw} S\law S$.

The identity map on objects and the map $$ R\stackrel{\rh}{\raw} S\:\:\:
\longmapsto \:\:\: R\stackrel{\rh}{\raw} S\law S $$ on arrows is a
contravariant functor from the category of rings with homomorphisms as
arrows into the bicategory \Ri.
\end{remark}

See \cite{Ben} for the notion of a functor between two bicategories;
we here look at the domain of the embedding functor as a 2-category
(cf.\ the preceding section).
The direction of arrows is reversed, since in \Ri, a bimodule $R\raw
M\law S$ is an arrow from $S$ to $R$.  This could be remedied by
mapping a homomorphism $\rh:R\raw S$ to the bimodule $S\raw
S\stackrel{\rh}{\law}R$, but this construction does not work for \ca s.

Morita's theorems give a necessary and sufficient condition for the
representation categories of two rings to be equivalent.
\begin{definition}\label{kejo}
The representation category $\Rep(R)$ of a ring $R$ has left
$R$-modules as objects, and $R$-module maps as arrows (that is, given
two $R$-modules $M$, $N$, elements of $(N,M)$ of arrows are
homomorphisms  from $M$ to $N$ as abelian groups that commute with
the $R$-action).
\end{definition}

In the present language, Morita theory starts as
follows.
\begin{theorem}\label{morita}
Two rings are isomorphic objects in the bicategory \Ri\ iff they have
equivalent \rep\ categories (where the equivalence functor is required
to be additive). 
\end{theorem}
\begin{remark} \label{Minverse}
The first  condition is explained in Definition \ref{eqbi}. To be
concrete, it here means that there exists an arrow in $(R,S)$ (that
is, a bimodule $R\raw M\law S$) that is invertible up to isomorphism.
Such an $M$ is often called an $R$-$S$ equivalence bimodule. In
other words, there should, in addition, exist an arrow $S\raw
M\inv\law R$ in $(S,R)$, such that
\begin{eqnarray*}
S\raw M\inv \ot_R M\law S &\cong & S\raw S\law S;  \\
R\raw M\ot_S M\inv \law R & \cong & R\raw R\law R. 
\end{eqnarray*}
Here the first $\cong$ stands for isomorphism in the category $(S,S)$,
whereas the second denotes isomorphism in $(R,R)$. The first condition
simply means that the composition of the two given bimodules, seen as arrows,
is isomorphic to the unit arrow in $(S,S)$, etc.
\end{remark}
\begin{proof}
The idea of the proof in the ``$\Raw$'' direction is as follows.
One constructs a functor $G:\Rep(S)\raw\Rep(R)$ by taking tensor
products: on objects one has $G_0(L)=M\ot_S L\in\Rep(R)_0$ for $L\in
\Rep(S)_0$, and on arrows one puts, in obvious notation,
$G_1(f)={\rm id}\ot_S f$. To go in the opposite direction, one
repeats the above procedure, in defining
a functor $F:\Rep(R)\raw\Rep(S)$ by means of
$F_0(N)=M\inv \ot_R N$, etc. Using Theorem \ref{biring}
and Remark \ref{Minverse}, 
it easily follows that $FG\simeq \id_{\Rep(S)}$ and 
 $GF\simeq \id_{\Rep(R)}$. 

In the ``$\Law$'' direction, one constructs $M$, given an equivalence
functor $G:\Rep(S)\raw\Rep(R)$, by putting $M=G_0(1_S)$.  The left $S$
action on $S$ is turned into a left $R$ action on $M$ by definition of
$G_0$, and the right $S$ action on $S$ is turned into a right $S$
action on $M$ through $G_1$, since $S\op\subseteq (S,S)\subset
\Rep(S)_1$. Thus $M\in (R,S)$. Similarly, define $F_0(1_R)\in
(S,R)$. The definition of equivalence of categories then trivially
implies that the isomorphisms in Remark
\ref{Minverse} hold, with $M\inv=F_0(1_R)$. 
For details, cf.\ no.\ 12.13 in \cite{faith}.
\end{proof}

The first part of this proof generalizes to all other classes of
mathematical objects we study in this paper. The second part, on the
other hand, only generalizes when the analogues of the identity arrows
$1_R$ lie in the \rep\ category under consideration, and when there is
enough functoriality around to turn the analogues of $F_0(1_R)$ into a
bimodule of the desired sort.  These two conditions are met in the
case of \vna s; in all other cases one has to adapt the setting.  In
this light, the following comment is relevant.
\begin{remark}\label{natrings}
An equivalence functor $F:\Rep(R)\raw\Rep(S)$ is automatically
fibered, in the following sense. For each fixed ring $T$, the functor
$F$ defines an equivalence $F_T$ between the categories $(R,T)$ and
$(S,T)$, natural in $T$.
\end{remark}

Naturality here means that, for any rings $T,T'$ and homomorphisms
$\phv:T\raw T'$, one has $\phv_S^* F_{T'}=F_T\phv_R^*$, where
$\phv^*_R:(R,T')\raw (R,T)$ is the induced functor.

In any case, the connection with the usual Morita theory is
\begin{proposition}\label{Minvt}
A bimodule  $M\in (R,S)$ is invertible as an arrow in \Ri\ iff
\begin{enumerate}
\item $R\simeq \End_{S^{\mbox{\tiny op}}}(M)$;
\item $M$ is finitely generated projective as an $R$- and as an 
$S^{\mbox{\tiny op}}$-module.
\end{enumerate}
In that case, the inverse may be taken as 
$M\inv=\Hom_{S^{\mbox{\tiny op}}}(M,S)$.
\end{proposition}
\begin{proof}
The  ``$\Raw$'' claim is part of ``Morita I'', cf.\ no.\ 12.10.4 
in \cite{faith} for 1, and 12.10.2  for 2. The converse
follows from nos.\  12.8(c) and 4.3(c) in \cite{faith}.
\end{proof}
\section{\ca s}
The $C^*$-algebraic analogue of
a bimodule for rings is a  Hilbert bimodule. This concept involves the theory
of Hilbert $C^*$-modules, for which we refer to \cite{RW,NPL3}. 
\begin{definition}\label{defhbm}
An $\A$-$\B$ Hilbert bimodule, where $\A$ and $\B$ are \ca s, is 
 a Hilbert $C^*$ module $\CE$ over $\B$, along with a nondegenerate
$\mbox{}^*$-homomorphism of $\A$ into $\CL_{\B}(\CE)$. We write
$\A\raw\CE\rlh\B$.
\end{definition}

The following example is the \ca ic version of the ring bimodule
$R\raw R\law R$.
\begin{example} \label{BB} A \ca\
$\B$ may be seen as a Hilbert bimodule $\B\raw\B\rlh\B$ over itself,
in which $\langle A,B\ra_{\B} =A^*B$, and the left and right actions
are given by left and right multiplication, respectively. 
\end{example}
 
Note that the norm in $\B$ as a \ca\ coincides with
its norm as a Hilbert module because of the $C^*$-axiom $\n A^*A\n=\n
A\n^2$.

The \ca ic analogue of the bimodule tensor product is
 the (interior) tensor product $\hat{\ot}_{\B}$ defined by
 Rieffel \cite{Rie}; also see \cite{RW,NPL3}.
In complete parallel with ring theory (cf.\ Theorem \ref{biring}),
one now has
\begin{theorem}\label{bicstar}
For any two \ca s, let $(\A,\B)$ be the collection of all Hilbert
bimodules $\A\raw \CE\rlh \B$, seen as a category, whose arrows are
adjointable $\A$ linear maps (such maps are automatically bimodule
maps).

With (horizontal) composition
$(\A,\B)\x (\B,\GC)\raw (\A,\GC)$ given by $\hat{\ot}_{\B}$, and unit
arrow in $ (\B,\B)$ given by $1_{\B}=\B\raw \B\rlh\B$, the collection
of all \ca s as objects, and Hilbert bimodules as arrows, forms a
bicategory \Ca.
\end{theorem}

Along the lines of Remark \ref{ringmor}, we have
\begin{remark}\label{cstamor}
An $\A$-$\B$ Hilbert bimodule may be seen as a generalization of
a nondegenerate $\mbox{}^*$-homomorphism $\rh:\A\raw\B$, for given such a $\rh$
one constructs a Hilbert bimodule $\A\raw\B\rlh\B$ by
$A(B)=\rh(A)B$, and the other operations as in Example \ref{BB}.
We write $\A\stackrel{\rh}{\raw} \B\rlh\B$.

Thus one obtains a contravariant functor from the category of \ca s
with $\mbox{}^*$-homomorphism as arrows into the bicategory \Ca.
\end{remark}

Rieffel, who launched the theory of Morita
equivalence of \ca s \cite{Rie}, defined the representation category
$\Rep(\A)$ of a \ca\ $\A$ as follows.
\begin{definition}\label{repsca}
The \rep\ category $\Rep(\A)$ of a \ca\ $\A$ has 
 nondegenerate \rep s of $\A$ on a Hilbert space as objects, and bounded linear
intertwiners as arrows.
\end{definition}

The first attempt to adapt  Theorem \ref{morita} to \ca s now reads
\begin{theorem}\label{rit}
If two \ca s are isomorphic
objects in the bicategory \Ca,
then they have equivalent \rep\ categories (in the above sense),
where the equivalence functor is
required to be linear and $\mbox{}^*$-preserving on arrows.
\end{theorem}

The proof is the same as for rings, the bicategory \Ca\ replacing \Ri.
Moreover, one should regard a \rep\ of a \ca\ $\A$ on a \Hs\ as an
$\A$-$\C$ Hilbert bimodule.

The \ca ic version of Proposition \ref{Minvt} is as follows.
\begin{proposition}\label{smeca}
A bimodule  $\CM\in (\A,\B)$ is invertible as an arrow in \Ca,
so that $\A$ and $\B$ are isomorphic in \Ca, iff
\begin{enumerate}
\item the linear span of the range of $\la \, ,\,\bb$ is dense in $\B$
(in other words, $\CM\rlh\B$ is full);
\item
the $\mbox{}^*$-homomorphism of $\A$ into $\CL_{\B}(\CE)$ of
Definition \ref{defhbm} is an isomorphism $\A\simeq\CK_{\B}(\CM)$.
(If $\A$ has a unit, this isomorphism will be $\A\simeq\CL_{\B}(\CM)$.)
\end{enumerate}

If $\A\raw\CM\rlh\B$ is invertible, its inverse (up to isomorphism) is
$\B\raw\ovl{\CM}\rlh \A$, where $\ovl{\CM}$ is the
conjugate space of $\CM$, on which $\B$ acts from the left by
$B:\Ps\mapsto B^*\Ps$ and $\A$ acts from the right by
$A:\Ps\mapsto A^*\Ps$. The $\A$-valued inner product on
$\ovl{\CM}$ is given by $\la\Ps,\Ph\ra_{\A}
=\phv\inv(T^{\B}_{\Ps,\Ph})$, where $\phv:\A\raw\CK_{\B}(\CM)$ is the
pertinent isomorphism.
\end{proposition}

Here $\CK_{\B}(\CM)$ is the \ca\ of ``compact'' operators on 
$\CM$, seen as a Hilbert $C^*$-module over $\B$
\cite{Rie,RW,NPL3}. Equivalent conditions are given in \cite{RW}.
\begin{proof}
This is essentially Prop.\ 2.3 in \cite{Sch} (Schweizer works with the
category of \ca s with equivalence classes of Hilbert bimodules as
arrows, rather than with the bicategory whose arrows are the Hilbert
bimodules themselves, but his proof may trivially be adapted to our
situation). We are indebted to Paul Muhly for drawing our attention
to this result. Muhly in addition provided a second proof in case
that the algebras are unital:
combine Thm.\ 4.3 in \cite{Ble} with Thm.\ 6.2 in \cite{BMP}.
\end{proof}

It is clear that the ring-theoretic proof of a potential ``$\Law$''
part of Theorem \ref{rit} cannot immediately be adapted to the present
case, since the bimodule $1_{\A}$ is not itself an element of
$\Rep(\A)$.  To remedy this defect, one should enlarge $\Rep(\A)$ so
that it contains $\A$.  This has been done by Blecher \cite{Ble2} in
the setting of operator spaces, operator modules, and completely
bounded maps, but here we choose a different way out, suggested by
Remark \ref{natrings}, and by a discussion with J. Mr\v{c}un.
\begin{theorem}\label{corblech}
Two \ca s $\A,\B$ are isomorphic in \Ca\ 
iff for any \ca\ $\GC$ one has a category equivalence $(\A,\GC)\simeq
(\B,\GC)$, natural in $\GC$.
\end{theorem}

The particular enlargement of $\Rep(\A)$ therefore involves all \rep s
of $\A$ on Hilbert modules over arbitrary \ca s $\GC$. This theorem
may be derived from the results in \cite{Ble2}; it is an open question
whether it can be proved by pure
\ca ic methods.
\section{von Neumann algebras}
Although one could adapt the theory of Hilbert bimodules for \ca s
so as to include normality of the actions, as in \cite{Rie},
there is a much simpler approach to bimodules for \vna s, initiated by
Connes \cite{Con}. 
\begin{definition}\label{defcor}
Let $\GM,\GN$ be \vna s. An $\GM$-$\GN$ correspondence
$\GM\raw\H\law\GN$ is given by a \Hs\ $\H$, a normal unital \rep\
$\pi$ of $\GM$ on $\H$, and a normal unital anti-\rep\ $\phv$ of $\GN$
on $\H$ (in other words, a normal \rep\ of $\GN\op$), such that
$\pi(\GM)\subseteq \phv(\GN)'$ (and hence $\phv(\GN)\subseteq
\pi(\GM)'$).
The correspondence is called faithful when $\pi$ and $\phv$ are injective.
\end{definition}

Matched correspondences may be composed using Connes's tensor
product  $\boxtimes_{\GN}$ \cite{Sau,Con}.
\begin{theorem}\label{bivna}
For any two \vna s, let $(\GM,\GN)$ be the collection of all 
correspondences $\GM\raw\H\law\GN$, seen as the object space of
 a category, whose arrows are bounded linear  bimodule maps.

The collection of all \vna s as objects, and correspondences as arrows,
forms a bicategory \Wa\ under (horizontal)  composition 
 $\boxtimes_{\GN}:(\GM,\GN)\x (\GN,\GP)\raw
(\GM,\GP)$, and the unit arrow in $(\GN,\GN)$  given by the 
standard form $\GN\raw L^2(\GN)\law\GN$.
\end{theorem}

The \vna ic version of Remarks \ref{ringmor} and \ref{cstamor}
holds verbatim \cite{Con}.

The theory of Morita equivalence of \vna s was initiated by Rieffel
\cite{Rie}, whose definition of strong Morita equivalence
was directly adapted from the \ca ic Proposition \ref{smeca}.  However,
the theory of correspondences enables one to rewrite his theory in a
way that practically copies the purely algebraic case of rings.
In any case, one puts \cite{Rie}
\begin{definition} \label{repvn}
Let $\GM$ be a \vna. The \rep\ category $\Rep(\GM)$ has normal unital \rep s
on \Hs s as objects, and boun\-ded linear intertwiners as arrows.
\end{definition}

The Morita theorem for \vna s then reads as follows:
\begin{theorem}
Two \vna s are  isomorphic objects in the bicategory \Wa\ iff their
\rep\ categories are equivalent (and the equivalence functor
implementing $\simeq$ is linear and $\mbox{}^*$-preserving on arrows).
\end{theorem}

The proof is the same as for rings, replacing the bicategory \Ri\ by
\Wa.  Von Neumann's bicommutant theorem provides enough functoriality
to make all steps of the ring proof to go through for \vna s.

Rieffel's original Morita theorem for \vna s was based on his own
definition of strong Morita equivalence (Def.\ 7.5 in \cite{Rie}),
which is equivalent to isomorphism in \Wa\ 
 by the following analogue of Propositions
\ref{Minvt} and \ref{smeca}:
\begin{proposition}\label{Minvna}
Two \vna s $\GM,\GN$ are isomorphic in \Wa\  iff
there exists a faithful correspondence $\GM\raw\H\law\GN$ for which
$\GM'\simeq\GN\op$ (and hence $(\GN\op)'\simeq\GM$).

In that case, the inverse correspondence is $\GN\raw\ovl{\H}\law\GM$,
where $\ovl{\H}$ is the conjugate of $\H$, and left and right actions 
are swapped using the involution on $\GN$ and $\GM$.
\end{proposition}
\begin{proof}
This follows from a combination of Theorem \ref{bicstar} above with
Thms.\ 7.9 and 8.15  in \cite{Rie} and Theorem 
 2.2 in \cite{BDH}. A direct proof would be desirable.
\end{proof}

The \vna ic version of Remark \ref{natrings} holds verbatim.
\section{Lie groupoids}
As for general categories, our generic notation for groupoids is that
$G_0$ is the base space of a groupoid $G$, with source and target maps
$s,t:G_1\raw G_0$, multiplication $m:G_2\raw G_1$ (where $G_2= G_1
*^{s,t}_{G_0}G_1$), inversion $I:G_1\raw G_1$, and object inclusion
$\io:G_0\hookrightarrow G_1$ (this inclusion map will often be taken
for granted, in that $G_0$ is seen as a subspace of $G_1$).

A Lie groupoid is a groupoid for which $G_1$ and $G_0$ are manifolds,
$s$ and $t$ are surjective submersions, and $m$ and $I$ are smooth.
It follows that $\io$ is an immersion, that $I$ is a diffeomorphism,
that $G_2$ is a closed submanifold of $G_1\x G_1$, and that for each
$q\in G_0$ the fibers $s\inv(q)$ and $t\inv(q)$ are submanifolds of
$G_1$.  References on Lie groupoids that are relevant to the themes in
this paper include \cite{Mac,CDW,NPL3}.

The following notion will be important in what follows: A Lie groupoid
(or, more generally, a topological groupoid) is called s-(simply) connected
 if the fibers of $s:G_1\raw G_0$ are (simply)
connected. 

We now define actions \cite{Mac} and bimodules \cite{Hae,HS,MRW,Moe,Mrc1,Mrc2}
for Lie groupoids. 
\begin{definition}\label{Gaction} 
\begin{enumerate}
\item
Let $G$ be a Lie groupoid and let $M\stackrel{\ta}{\raw} G_0$ be smooth.
  A left $G$ action on $M$ (more precisely, on $\ta$) is a smooth
map $(x,m)\mapsto xm$ from $G *^{s,\ta}_{G_0}M$ to $M$
(i.e., one has $s(x)=\ta(m)$), such
that $\ta(xm)=t(x)$, $xm=m$ for all $x\in G_0$, and $x(ym)=(xy)m$
whenever $s(y)=\ta(m)$ and $t(y)=s(x)$.  
\item
A right action of a Lie groupoid $H$ on $M\stackrel{\sg}{\raw} H_0$ is
a smooth map $(m,h)\mapsto mh$ from $M*^{t,\ta}_{H_0} H$ to $M$ that
satisfies $\sg(mh)=s(h)$, $mh=m$ for all $h\in H_0$, and 
$(mh)k=m(hk)$ whenever $\sg(m)=t(h)$ and $t(k)=s(h)$.
\item
A $G$-$H$ bibundle $M$  carries a left
$G$ action as well as a right $H$ action that commute. That is,
 one has
$\ta(mh)=\ta(m)$, $\sg(xm)=\sg(m)$, and $(xm)h=x(mh)$ for all $(m,h)\in
M*H$ and $(x,m)\in G*M$. On occasion, we simply write $G\raw M\law H$.
\end{enumerate}
The maps $\ta$ and $\sg$ will sometimes be called the base maps of the
given actions.
\end{definition}

In the purely algebraic case, one may form a tensor product between
two matched groupoid bibundles $G\raw M\law H$ and $H\raw N\law K$, as
follows.  The pullback $M*_H N$ carries a diagonal right $H$ action, given by
$h:(m,n)\mapsto (mh,h\inv n)$ (defined as appropriate). The tensor
product over $H$ is then simply the orbit space $$ M\circledast_H
N=(M*_H N)/H, $$ seen as a $G$-$K$ bibundle under the obvious maps.
We name this tensor product after Hilsum--Skandalis \cite{HS}.
In the smooth case, one needs  further assumptions for this construction
to work.
\begin{definition}\label{defprin}
A (left)  $G$ bundle $M$ over a manifold $X$ consists of
a (left) $G$ action on $M$ and a smooth map $\pi:M\raw X$ that
is invariant under the $G$ action. Similarly for right actions.

A (left) $G$ bundle $M$ over $X$ is called principal when $\pi$ is a
surjective submersion and the map from $G_1 *_{G_0}^{s,\ta}M\raw M*_X
M$ given by $(x,m)\mapsto (xm,m)$ is a diffeomorphism.  In other
words, the action is free (in that $xm=m$ iff $x\in G_0$) and
transitive along the fibers of $\pi$, and one has $G\backslash M\simeq
X$ through $\pi$.

 A $G$-$H$ bibundle $M$ is called left principal when it is principal for the
 $G$ action with respect to $X=H_0$ and $\pi=\sg$.
Similarly, it is called right principal when it is principal for the
 $H$ action with respect to $X=G_0$ and $\pi=\ta$.

A $G$-$H$ bibundle $M$ is called regular when it is left principal
and the right $H$ action is proper (in that the map $(m,h)\mapsto
(m,mh)$ from $M*_{H_0}H$ to $M\x M$ is proper).
\end{definition}

The definition of a principal bundle is taken from \cite{Hae,Mrc1,Mrc2};
it is different from the one in \cite{MRW}. 

Now, if two Lie groupoid bibundles $G\raw M\law H$ and $H\raw N\law K$
are both regular, then the tensor
product $M\circledast_H N$ is a manifold, and is even a $G$-$K$
bibundle. For the surjectivity of $\sg:M\raw H_0$ 
implies that $M*_{H_0} N$ is a submanifold of $M\x N$, and the freeness
of the $H$ action on $N$ with the properness of the $H$ action
on $M$ implies that the diagonal $H$ action on $M*_{H_0}N$ is free and
proper, so that  the quotient space
$M\circledast_H N$ is a manifold.

To explain the unit objects in the bicategory \LG\ to be defined
shortly,  first note the following \cite{Mrc1,Mrc2}.
\begin{remark}\label{smbi}
A homomorphism $\Ps:G\raw H$ is a smooth functor (in that
$\Ps_0$ and $\Ps_1$ are smooth).
Such a functor  defines a $G$-$H$ bibundle with total
space $M=G_0*^{\Ps_0,t}_{H_0} H$, base maps $pr_1:M\raw G_0$ and
$s_2:M\raw H_0$, a left $G$ action inherited from the obvious
$G$ action $G*^{s,\mbox{\rm id}}_{G_0} G_0\raw G_0$, and the right $H$ action
is given by multiplication.
Here $pr_1$ is projection onto the first coordinate, and $s_2$
is essentially the source projection of $H$.
\end{remark}

A special case is provided by $H=G$ and the identity functor $\Ps={\rm
id}:G\raw G$, leading to the canonical $G$-$G$ bibundle $G$.
This arises from the general case  by the isomorphism
$G*^{t,\mbox{\rm id}}_{G_0} G_0\simeq G$, $(x,q)\mapsto x$, and may be seen
as the groupoid version $1_G$ of the ring bimodule $R\raw R\law R$,
or of the standard form of a \vna.

Thus one obtains a version of Theorem \ref{biring} for Lie groupoids:
\begin{theorem}\label{bigroupoids}
For any two Lie groupoids, let $(G,H)$ be the collection of all
regular $G$-$H$ bibundles, seen as the object
space of a category, in which an arrow from $M$ to $N$ is a smooth map
that intertwines the maps $M\raw G_0$, $M\raw H_0$ with the maps
$N\raw G_0$, $N\raw H_0$, and in addition intertwines the $G$ and $H$
actions (the latter condition is well defined because of the former).

Then the collection of all Lie groupoids as objects, regular bibundles
as arrows, (horizontal) composition $(G,H)\x (H,K)\raw (G,K)$ given by
the Hilsum--Skandalis tensor product $\circledast_H$, and the unit
arrow $1_H$ in $(H,H)$ as defined above, is a bicategory \LG.
\end{theorem}

This theorem was inspired by \cite{Mrc1}, where the case of
topological groupoids is discussed in the setting of ordinary
categories, so that one works with equivalence classes of the
bimodules in question. Dropping our right properness assumption, such
classes are sometimes called Hilsum--Skandalis maps, following their
introduction in the special case of groupoids defined by foliations
\cite{HS}. Also cf.\ \cite{Hae,Moe,Mrc2}.

Remark \ref{smbi} leads to a satisfactory groupoid counterpart of
Remarks \ref{ringmor} and \ref{cstamor}, which we leave to the reader
to spell out.

We now write down the Lie groupoid counterpart
of Definition \ref{kejo}.
\begin{definition}\label{Repgoid}
The objects of the \rep\ category $\Rep(G)$ of a Lie group\-o\-id
$G$ are left $G$ actions on smooth maps $M\stackrel{\ta}{\raw} G_0$.
The space of arrows $(N,M)$ between a \rep\ on $M\stackrel{\ta}{\raw} G_0$
and one on $N\stackrel{\sg}{\raw} G_0$ consists of all smooth maps
$\phv:M\raw N$ that satisfy $\sg\phv=\ta$ and intertwine the $G$ action.
\end{definition}

As in Theorem \ref{morita}, we now  have
\begin{theorem} \label{Morgr}
If two Lie groupoids are isomorphic objects in the bicategory \LG,
then their \rep\ categories are equivalent.
\end{theorem}
The proof is literally the same as for rings, the bicategory \LG\
replacing \Ri.

As pointed out by I. Moerdijk, a ``$\LRaw$'' Morita theorem is
possible along the lines of Remark \ref{natrings}. 
The equivalence functor $\Rep(G)\raw\Rep(H)$
provided by an isomorphism $G\cong H$ automatically
induces equivalence functors $(G,K)\raw (H,K)$, natural in
$K$. Unlike the case of rings and \vna s, this property is not shared
by arbitrary equivalence functors, which is the reason why there is
no ``$\LRaw$'' implication in Theorem \ref{Morgr}. One then has the
following statement: two (Lie) groupoids $G,H$ are strongly Morita
equivalent iff for all $K$ there exist equivalences $(G,K)\simeq(H,K)$,
natural in $K$. (It is, in fact, only necessary to have such
equivalences for a small class of $G$-$K$ bibundles, namely those
in which $K$ is the trivial groupoid over itself, acting trivially
on the middle space.)
\begin{proposition}\label{MEGR}
Two Lie groupoids $G,H$ are isomorphic objects in the bicategory \LG\
iff there exists a $G$-$H$ bibundle $M$
with the additional properties:
\begin{enumerate}
\item $M$ is  left and right principal;
\item The $G$  and $H$ actions are proper.
\end{enumerate}

In that case, an inverse of $G\raw M\law H$ is 
the bibundle  $H\raw \ovl{M}\law G$, where $\ovl{M}$ is
$M$ as a manifold, seen as a $H$-$G$
bibundle with the same base maps, and left and right actions
interchanged using the inverse in $G$ and $H$.
\end{proposition}

This  reproduces the definition of Morita
equivalence for (Lie) groupoids given in \cite{MRW}.  Other 
definitions that are in use are equivalent to this one; 
see, e.g., \cite{Hae,Moe,Mrc2,Xu1},
and refs.\ therein. 
\begin{proof} 
The second part of the proposition, which at the same
time proves the ``$\Law$'' claim in the first part, is proved by the
argument following Def.\ 2.1 in \cite{MRW}.

For the ``$\Raw$'' claim in the first part, we are given a regular bibundle
$G\raw M\law H$, with regular inverse $H\raw M\inv\law G$.
This leads to two isomorphisms, as displayed in Remark \ref{Minverse}
for rings. From $M\circledast_H M\inv\simeq G$ as $G$-$G$ bibundles
we infer that the left $G$ action on $M\circledast_H M\inv$ is proper
(since the canonical left $G$ action on $G$ is), which
by reductio ad absurdum implies that the left $G$ action on $M$ is
proper. Since the target projection $t:G_1\raw G_0$ is a
surjective submersion, so is the map $M\circledast_H M\raw G_0$,
and therefore $\ta:M\raw G_0$ must be a surjective submersion as well.
In similar vein, the isomorphism $M\inv\circledast_G M\simeq H$ as $H$-$H$
 bibundles implies that the right $H$ action on $M$ is free and
transitive on the $\ta$-fibers. Together with the assumed
regularity of the bibundle $G\raw M\law H$, we have now proved 
both conditions in the proposition.
\end{proof} 
\section{Symplectic groupoids}
We now specialize to symplectic groupoids, which will be crucial for a
bicategorical understanding of Poisson manifolds, and are of interest
in themselves. Our basic references are \cite{MiWe,CDW}.
\begin{definition}\label{defsg}
A symplectic groupoid is a Lie groupoid $\Gm$ for which $\Gm_1$ is a symplectic
manifold, with the property that the graph of $\Gm_2\subset \Gm\x \Gm$
is a Lagrangian submanifold of $\Gm\x \Gm\x \Gm^-$.
\end{definition}

The notion of a bibundle for symplectic groupoids is an
adaptation of Definition \ref{Gaction}, now also involving the 
idea of a symplectic groupoid action \cite{MiWe}.
\begin{definition}\label{defsgaction}
An action of a symplectic groupoid $\Gm$ on a symplectic manifold
$S$ is called symplectic when the graph of the action in
$\Gm\x S\x S^-$ is Lagrangian.

Let $\Gm,\Sg$ be symplectic groupoids. A (regular) symplectic $\Gm$-$\Sg$
bibundle consists of a symplectic space $S$ that is a (regular) bibundle as in
Definition \ref{Gaction}, with the additional requirement that the two
groupoid actions be symplectic.
\end{definition}

The tensor product of two matched regular bibundles for symplectic groupoids
is then defined exactly as in the general (non-symplectic) case.
Hence we obtain a symplectic version of Theorem \ref{bigroupoids}:
\begin{theorem}\label{sygroupoids}
For any two symplectic groupoids $\Gm,\Sg$, let $(\Gm,\Sg)$ be the
collection of all regular  symplectic $\Gm$-$\Sg$
bibundles, seen as the class of objects of a category, with arrows as
described in Proposition \ref{bigroupoids}, with the additional
requirement that the maps preserve the symplectic form.

Then the collection of all symplectic groupoids as objects, regular
 symplectic bibundles as arrows, 
(horizontal) composition $(\Gm,\Sg)\x (\Sg,\Upsilon)\raw (\Gm,\Upsilon)$ 
given by
$\circledast_{\Sg}$, and the unit arrow $1_{\Gm}$ in $(\Gm,\Gm)$ given by the
canonical bibundle over $\Gm$,   is a bicategory
\SyG.
\end{theorem}

The discussion of Morita equivalence for symplectic groupoid is an
obvious adaptation of the Lie groupoid case.
\begin{definition}\label{Repsg}
The objects of the \rep\ category $\Rep^s(\Gm)$ of a symplectic groupoid
$\Gm$ are symplectic left $\Gm$ actions on 
smooth maps $\ta:S\raw\Gm_0$, where $S$ is symplectic.
The space of arrows is as in Definition \ref{Repgoid}, with
the additional requirement that $\phv$ be a complete Poisson map
(cf.\ the next section).
\end{definition}

It should be mentioned that $\ta$ is necessarily a complete Poisson map
\cite{MiWe,CDW}.
As in Theorem \ref{Morgr}, we again have
\begin{theorem} \label{Morsg}
If two symplectic groupoids are isomorphic in \SyG, then their
\rep\ categories $\Rep^s(\cdot)$ are equivalent.
\end{theorem}

Xu's original
Morita theorem for symplectic groupoids \cite{Xu1}
is now equivalent to the following statement, which is proved like
Proposition \ref{MEGR}.
\begin{proposition}\label{isosg}
Two symplectic groupoids $\Gm,\Sg$ are  isomorphic in \SyG\
 iff as Lie groupoids they satisfy
the condition stated in Proposition \ref{MEGR}, with the additional
requirement that the middle space $M=S$ 
is symplectic, and  that the $\Gm$ and $\Sg$ actions
are symplectic (see Definition \ref{defsgaction}).

In that case, an inverse bibundle is $S^-$, seen as a $\Sg$-$\Gm$
bibundle by interchanging left and right actions through the groupoid inverse.
\end{proposition}
\section{Poisson manifolds}
 Poisson algebras are the
classical analogues of \ca s and \vna s; see, e.g.,
\cite{NPL3}.
\begin{definition}
A Poisson algebra is a commutative associative algebra $A$
(over $\C$ or $\mathbb R$) endowed with a Lie bracket $\{\, ,\, \}$
such that each $f\in A$ defines a derivation $X_f$ on $A$ (as a
commutative algebra) by $X_f(g)=\{f,g\}$. In other words, the Leibniz
rule $\{f,gh\}=\{f,g\} h+g\{f,h\}$ holds.

A Poisson manifold is a manifold $P$ with a Lie bracket on $\cin(P)$
such that the latter becomes a Poisson algebra under pointwise
multiplication. 
\end{definition}

We write $P^-$ for $P$ with minus a given Poisson bracket.  Not all
Poisson algebras are of the form $A=\cin(P)$ (think of singular
reduction), but we specialize to this case. The derivation $X_f$ then
corresponds to a vector field on $P$, called the Hamiltonian vector
field of $f$. If the span of all $X_f$ (at each point) is $TP$, then
$P$ is symplectic.

The classical counterpart of a Hilbert bimodule or a correspondence is
a symplectic bimodule. 
First, recall that a Poisson map $q:S\raw Q$ between Poisson manifolds
is a smooth map whose pullback $q^*:\cin(Q)\raw \cin(S)$ is a
homomorphism of Poisson algebra. A Poisson map $q$, or rather its
pullback $q^*$, defines  a pair of maps.
The first of these, $q^*_c$, defines
$\cin(S)$ as a module for $\cin(Q)$ as a commutative algebra
through $q^*_c(f)g= (q^*f)g$. The second, $q^*_L$,
maps $\cin(Q)$ into the Lie algebra $\Der(\cin(S))$ of derivations of $\cin(S)$
through $q^*_L(f)=X_{q^*f}$, or, in other words, $q^*_L(f)g=\{q^*f,g\}$.
This map is a Lie algebra homomorphism by definition of a Poisson map.

A Poisson map $q:S\raw Q$ is called complete when, for every
$f\in\cin(Q)$ with complete Hamiltonian flow, the Hamiltonian
flow of $q^*f$ on $S$ is complete as well (that is, defined for
all times). Requiring a Poisson map to be complete is a classical
analogue of the self-adjointness condition on a \rep\ of a \ca\ 
\cite{NPL3}. The following definition goes back to Weinstein
 \cite{W1} and Karasev \cite{K1}.
\begin{definition}\label{defsb}
A  symplectic bimodule $Q\law S \raw P$ consists of a
symplectic manifold $S$, Poisson manifolds $Q$ and $P$, and complete Poisson
maps $q:S\raw Q$ and $p:S\raw P^-$, such
that $\{q^* f, p^* g\}=0$ for all $f\in\cin(Q)$ and $g\in\cin(P)$.
\end{definition}

In order to give a bicategorical description of Poisson manifolds,
we need to explain the connection between Poisson manifolds and
symplectic groupoids.
The compatibility condition in Definition \ref{defsg}
expresses the idea that groupoid multiplication
should be a Poisson map. This has the following consequence
\cite{CDW,MiWe}.
\begin{proposition}\label{sgbasic}
Let $\Gm$ be a symplectic groupoid.
There exists a unique Poisson structure on $\Gm_0$ such that
$t$ is a complete Poisson map and $s$ is a complete anti-Poisson map.
Hence $\Gm_0\stackrel{t}{\law} \Gm\stackrel{s}{\raw} \Gm_0$
is a symplectic bimodule.
\end{proposition}

The objects in the bicategory \Po\ will be of the following form 
\cite{CDW,NPLOBWF}.
\begin{definition}\label{defintP}
A Poisson manifold $P$ is called integrable when there
exists an $s$-connected and $s$-simply connected
symplectic groupoid $\Gm(P)$ over $P$ (so that $P$ is isomorphic
to $\Gm(P)_0$ as a Poisson manifold).
\end{definition}

It can be shown that $\Gm(P)$ is unique up to isomorphism 
(see Thm.\ 3.21 in \cite{NPLOBWF}).
Running ahead of the Morita theory for Poisson manifolds, we here need
\begin{definition}
The \rep\ category $\Rep(P)$ of a Poisson manifold has complete
Poisson maps $q:S\raw P$, where $S$ is some symplectic space, as
objects, and complete Poisson maps $\phv:S_1\raw S_2$, where
$q_2\phv=q_1$, as arrows.
\end{definition}

This definition goes back to Weinstein (see, e.g., \cite{W1,CDW}), although
we use a more straightforward choice of arrows. In any case, one now
has the following extraordinary relationship between 
Poisson manifolds and symplectic groupoids.
\begin{theorem}\label{xd}
If $\Gm(P)$ and $P$ are related as in Definition \ref{defintP}, then
the categories $\Rep^s(\Gm(P))$ and $\Rep(P)$ are equivalent.
\end{theorem}

This theorem is independently due to Dazord and Xu \cite{Daz,Xu2}; 
key ingredients of the proof already appeared in \cite{CDW,MiWe}.
\begin{corollary}\label{ctplemma}
\begin{enumerate}
\item
Let $P$ and $Q$ be integrable Poisson manifolds, with associated
s-connected and s-simply connected symplectic groupoids $\Gm(P)$ and
$\Gm(Q)$; cf.\ Definition
\ref{defintP}. 

There is a natural bijective correspondence between
symplectic bimodules $Q\law S\raw P$ and symplectic bibundles
$\Gm(Q)\raw S \law\Gm(P)$.
\item
Let $R$ be a third integrable Poisson manifold, with associated
s-connected and s-simply connected symplectic groupoid $\Gm(R)$,
and let $Q\law S_1\raw P$ and 
$P\law S_2\raw R$ be symplectic bimodules. 

The Hilsum--Skandalis tensor product $S_1\circledast_{\Gm(P)} S_2$ of the
associated symplectic bibundles $\Gm(Q)\raw S_1\law\Gm(P)$ and
$\Gm(P)\raw S_2\law\Gm(R)$ is a $Q$-$R$ symplectic bimodule.
\end{enumerate}
\end{corollary}

This follows from Theorem \ref{xd}. In the setting of Poisson
manifolds and symplectic groupoids, the Hilsum--Skandalis tensor
product was introduced by Xu \cite{Xu1}; also cf.\ \cite{MiWe}. It may
alternatively be described without groupoids using the special
symplectic reduction procedure of \cite{NPL3}, in which case we write
it as $S_1\otc_P S_2$ ($=S_1\circledast_{\Gm(P)} S_2$).  This remark
provides the connection between the mathematical structures in this
paper and the physics of constrained dynamical systems.

In the following theorem, given two $Q$-$P$ symplectic bimodules
$Q\stackrel{q_i}{\law}S_i\stackrel{p_i}{\raw}P$ for $i=1,2$ we say
that a smooth map $\phv:S_1\raw S_2$ is a $Q$-$P$ map when
$q_2\phv=q_1$ and $p_2\phv=p_1$. Also, we call a $Q$-$P$ symplectic 
bimodule regular when the associated $\Gm(Q)$-$\Gm(P)$ symplectic
bibundle is regular (cf.\ Corollary \ref{ctplemma}.1).
\begin{theorem}
For any two integrable Poisson manifolds, let $(P,Q)$ consist
of all regular symplectic bimodules $P\law S\raw Q$. 
The class $(Q,P)$ consists of the objects of a category, whose arrows
are complete $Q$-$P$ Poisson maps 

Then the class of all integrable Poisson manifolds as objects,
regular symplectic bimodules as arrows, (horizontal)
composition of arrows $(Q,P)\x (P,R)\raw (Q,R)$ given by the
Hilsum--Skandalis--Xu tensor product as in Corollary \ref{ctplemma}, and
the unit arrow $1_P\in (P,P)$ given by $\Gm(P)$, forms a bicategory
\Po.
\end{theorem}
\begin{proof}
This follows from Theorem \ref{sygroupoids} and Corollary \ref{ctplemma}.
\end{proof}

It would be desirable to translate the stated conditions on the $\Gm(P)$ and
$\Gm(Q)$ actions into conditions on the maps $p$ and $q$.

The Poisson analogue of Remarks \ref{ringmor}, \ref{cstamor}, etc.\ is
as follows:
\begin{remark}\label{frctp}
A complete Poisson map $\rh:P\raw Q$ defines a symplectic bimodule 
$Q\stackrel{\rh}{\law}\Gm(P)\raw P$, so that a symplectic bimodule
may be seen as a generalized Poisson map.
The identity map on objects and the map $$ 
P\stackrel{\rh}{\raw} Q\:\:\:
\longmapsto \:\:\: Q\stackrel{\rh}{\law}\Gm(P)\raw P $$ on arrows is a
covariant functor from the category of Poisson manifolds
 with Poisson maps as
arrows into the bicategory \Po.
\end{remark}

The theory of Morita equivalence of Poisson manifolds was initiated by
Xu \cite{Xu2}, and may now be reformulated as follows. 
\begin{theorem}\label{xuthm}
If two integrable Poisson manifolds are isomorphic objects in the
bicategory \Po, then their \rep\ categories are equivalent.
\end{theorem}

This is proved as for rings. Interestingly, one has \cite{Xu2,NPLOBWF}
\begin{proposition}\label{inteq}
 A Poisson manifold $P$ is  integrable iff it is Morita equivalent to
itself.
\end{proposition}

Xu's original definition of \Me\ of Poisson manifolds 
\cite{Xu2} now becomes an inference.
\begin{proposition}\label{MEPM}
The following statements are equivalent:
\begin{enumerate}
\item
Two integrable Poisson manifolds $P,Q$ are  isomorphic objects in
\Po.
\item
There is a symplectic bimodule $Q\law S\raw P$ with the following
additional properties:
\begin{enumerate}
\item
The maps $p:S\raw P$ and $q:S\raw Q$ are surjective submersions;
\item The level sets of $p$ and $q$ are connected and simply connected;
\item The foliations of $S$ defined by the
levels of $p$ and $q$ are mutually symplectically orthogonal (in that
the tangent bundles to these foliations are each other's
  symplectic orthogonal complement).
\end{enumerate}
\item
The symplectic groupoids $\Gm(P)$ and $\Gm(Q)$ are \Me.
\end{enumerate}

In that case, an inverse of $Q\law S\raw P$ is $P\law S^-\raw Q$,
with the same maps.
\end{proposition}

\begin{proof}
The equivalence of 2 and 3 is Thm.\ 3.2 in \cite{Xu2}. 

By Theorem \ref{isosg}, statement 3 is equivalent to 
$\Gm(Q)\cong\Gm(P)$ in \SyG, 
which by Definition \ref{eqbi} means that there is an invertible
bibundle $\Gm(Q)\raw S\law\Gm(P)$ in \SyG.
By Corollary \ref{ctplemma} this is equivalent to
the invertibility of the associated symplectic bimodule
$Q\law S\raw P$ in \Po, which by Definition \ref{eqbi} is statement 1.
\end{proof}

\end{document}